\documentstyle[preprint,tighten,aps]{revtex}
\begin{document}
\title{  Proton drip-line nuclei in 
Relativistic Hartree-Bogoliubov theory}
\author{D. Vretenar$^1$, G.A. Lalazissis$^2$, 
and P. Ring$^2$}
\address{$^{1}$ Physics Department, Faculty of Science, University of
Zagreb, Croatia}
\address{$^{2}$ Physik-Department der Technischen Universit\"at M\"unchen, 
Garching, Germany}
\date{\today}
\maketitle
\begin{abstract}
Ground-state properties of spherical even-even nuclei
$14\leq Z \leq 28$ and $N=18,20,22$ are described in the framework  
of Relativistic Hartree Bogoliubov (RHB) theory. The model uses
the NL3 effective interaction in the mean-field Lagrangian,
and describes pairing correlations 
by the pairing part of the finite range Gogny interaction D1S. 
Binding energies, two-proton separation energies, and 
proton $rms$ radii that result from fully self-consistent
RHB solutions are compared with experimental data.
The model predicts the location of the proton drip-line.
The isospin dependence of the effective spin-orbit potential is 
discussed, as well as pairing properties that result from
the finite range interaction in the $pp$ channel.
\end{abstract}
\vspace{1 cm}
%
\section {Relativistic Hartree Bogoliubov theory with
finite range pairing interaction}
The structure of proton-rich nuclei presents many interesting phenomena 
which are very important both for nuclear physics and astrophysics.
These nuclei are characterized by exotic ground-state decay 
properties such as direct emission of charged particles and $\beta$-decays
with large Q-values. The properties of most proton-rich nuclei
should also play an important role in 
the process of nucleosynthesis by rapid-proton 
capture. In addition to decay properties (particle 
emission, $\beta$-decay), of fundamental importance are 
studies of atomic masses and separation energies, and especially the 
precise location of proton drip-lines. On the other hand, 
nuclear-structure models can be compared in detailed theoretical 
studies of nuclei with a large proton excess. In particular, 
for proton-rich nuclei in the $sd-f$ shell ($10\leq Z\leq 28$) 
predictions of the nuclear shell-model can be compared with
results of models that are based on the mean-field approach.
Shell-model calculations of proton-rich nuclei with 
$37\leq A\leq 48$ have recently been reported in Ref. \cite{Orm.96},
and the structure of proton-drip line around $^{48}$Ni has been
investigated in the framework of the self-consistent mean-field
theory in Ref. \cite{Naz.96}. In addition to non-relativistic
Hartree-Fock and Hartree-Fock-Bogoliubov models, in Ref. \cite{Naz.96}
also the relativistic mean-field model has been used, with pairing
properties described in the BCS approximation. While this approximation
is acceptable for nuclei close to the $\beta$-stability line, 
as we move away from the valley of $\beta$-stable nuclei the
ground-state properties calculated with the BCS scheme 
become unreliable. For proton-rich nuclei this problem 
might be less critical than for nuclei at the neutron drip-line, 
but nevertheless we expect a much better description of 
ground-state properties in a framework which provides 
a unified description of mean-field and pairing correlations.

In the present study we report the first application of 
Relativistic Hartree-Bogoliubov (RHB) theory to the structure
of proton-rich nuclei. Models based on quantum hadrodynamics
have been extensively applied in
calculations of nuclear matter and properties of finite
nuclei throughout the periodic table. In the self-consistent mean-field
approximation, detailed calculations have been performed
for a variety of nuclear structure phenomena \cite{Rin.96}. For open
shell nuclei pairing correlations have been included in the usual BCS 
approximation scheme, but also more consistently in the 
Hartree-Bogoliubov framework.
RHB presents a relativistic extension of the Hartree-Fock-Bogoliubov
theory. It was derived in Ref. \cite{KR1.91} in an attempt to
develop a unified framework in which  relativistic mean-field and
pairing correlations could be described simultaneously. As in 
ordinary HFB, the ground state of a nucleus $\vert \Phi >$ is described
as vacuum with respect to independent quasi-particle operators,
which are defined by a unitary Bogoliubov transformation of the
single-nucleon creation and annihilation operators.  The
generalized single-nucleon Hamiltonian 
contains two average potentials: the self-consistent mean-field
$\hat\Gamma$ which encloses all the long range {\it ph}
correlations, and a pairing field $\hat\Delta$ which sums
up the {\it pp} correlations. The expectation value of the
nuclear Hamiltonian (non-relativistic or Dirac) 
$< \Phi\vert \hat H \vert \Phi >$ is
a function of the hermitian density matrix
$\rho$, and the antisymmetric pairing tensor $\kappa$. The
variation of the energy functional with respect to $\rho$
and $\kappa$ produces the single quasi-particle 
Hartree-Fock-Bogoliubov equations~\cite{RS.80} in the
non-relativistic framework.
In the relativistic extension ~\cite{KR1.91}
the Hartree approximation is employed for
the self-consistent mean field, and the resulting Relativistic
Hartree-Bogoliubov (RHB) equations read
\begin{eqnarray}
\label{equ.2.2}
\left( \matrix{ \hat h_D -m- \lambda & \hat\Delta \cr
                -\hat\Delta^* & -\hat h_D + m +\lambda} \right) 
\left( \matrix{ U_k({\bf r}) \cr V_k({\bf r}) } \right) =
E_k\left( \matrix{ U_k({\bf r}) \cr V_k({\bf r}) } \right).
\end{eqnarray}
where $\hat h_D$ is the single-nucleon Dirac
Hamiltonian (\ref{statDirac}) and $m$ is the nucleon mass.
The chemical potential $\lambda$  has to be determined by
the particle number subsidiary condition, in order that the
expectation value of the particle number operator
in the ground state equals the number of nucleons. The column
vectors denote the quasi-particle wave functions, and $E_k$
are the quasi-particle energies. The Dirac Hamiltonian 
\begin{equation}
\label{statDirac}
\hat h_D~=~-i\mbox{\boldmath $\alpha$}
\cdot\mbox{\boldmath $\nabla$}
+\beta(m+g_\sigma \sigma)
+g_\omega \omega^0+g_\rho\tau_3\rho^0_3
+e\frac{(1-\tau_3)}{2} A^0
\end{equation}
contains the mean-field potentials of the isoscalar 
scalar $\sigma$-meson, the isoscalar vector $\omega$-meson
and the isovector vector $\rho$-meson. $A^0$ is the 
electrostatic potential. 
The RHB equations have to be solved self-consistently, with
potentials determined in the mean-field approximation from
solutions of Klein-Gordon equations
\begin{eqnarray}
\label{equ.2.3.a}
\bigl[-\Delta + m_{\sigma}^2\bigr]\,\sigma({\bf r})&=&
-g_{\sigma}\,\rho_s({\bf r})
-g_2\,\sigma^2({\bf r})-g_3\,\sigma^3({\bf r})   \\
\label{equ.2..3.b}
\bigl[-\Delta + m_{\omega}^2\bigr]\,\omega^0({\bf r})&=&
-g_{\omega}\,\rho_v({\bf r}) \\
\label{equ.2.3.c}
\bigl[-\Delta + m_{\rho}^2\bigr]\,\rho^0({\bf r})&=&
-g_{\rho}\,\rho_3({\bf r}) \\
\label{equ.2.3.d}
-\Delta \, A^0({\bf r})&=&e\,\rho_p({\bf r}).
\end{eqnarray}
for the sigma meson, omega meson, rho meson and photon
field, respectively. The spatial components 
\mbox{\boldmath $\omega,~\rho_3$} and ${\bf  A}$
vanish due to time reversal
symmetry. The equation for the sigma meson contains the 
non-linear $\sigma$ self-interaction terms~\cite{BB.77}.
Because of charge conservation, only the
3-component of the isovector rho meson contributes. The
source terms in equations (\ref{equ.2.3.a}) to
(\ref{equ.2.3.d}) are sums of bilinear products of baryon
amplitudes
\begin{eqnarray}
\label{equ.2.3.e}
\rho_s({\bf r})&=&\sum\limits_{E_k > 0} 
V_k^{\dagger}({\bf r})\gamma^0 V_k({\bf r}), \\
\label{equ.2.3.f}
\rho_v({\bf r})&=&\sum\limits_{E_k > 0} 
V_k^{\dagger}({\bf r}) V_k({\bf r}), \\
\label{equ.2.3.g}
\rho_3({\bf r})&=&\sum\limits_{E_k > 0} 
V_k^{\dagger}({\bf r})\tau_3 V_k({\bf r}), \\
\label{equ.2.3.h}
\rho_{\rm em}({\bf r})&=&\sum\limits_{E_k > 0} 
V_k^{\dagger}({\bf r}) {{1-\tau_3}\over 2} V_k({\bf r}).
\end{eqnarray}
where the sums run over all positive energy states. 
The pairing field
$\hat\Delta $ in (\ref{equ.2.2}) is defined
\begin{equation}
\label{equ.2.5}
\Delta_{ab} ({\bf r}, {\bf r}') = {1\over 2}\sum\limits_{c,d}
V_{abcd}({\bf r},{\bf r}') {\bf\kappa}_{cd}({\bf r},{\bf r}').
\end{equation}
where $a,b,c,d$ denote quantum numbers
that specify the single-nucleon states.
$V_{abcd}({\bf r},{\bf r}')$ are matrix elements of a
general two-body pairing interaction, and the pairing
tensor is defined 
\begin{equation}
{\bf\kappa}_{cd}({\bf r},{\bf r}') = 
\sum_{E_k>0} U_{ck}^*({\bf r})V_{dk}({\bf r}').
\end{equation}

In the Relativistic Hartree-Bogoliubov theory
pairing correlations result from the one-meson exchange
($\sigma$-, $\omega$- and $\rho$-mesons) \cite{KR1.91}. 
However, if for the pairing part of the interaction 
one uses the coupling 
constants from standard parameter sets of the 
relativistic mean-field model, 
the resulting pairing correlations are much too strong.
The repulsion produced by the 
exchange of vector mesons at short distances results in a 
pairing gap at the Fermi surface
that is by a factor three too large.
This is not surprising, since in general 
the effective interactions in the particle-hole and 
particle-particle channels do not have to be identical. 
In a first-order approximation,
the effective interaction contained 
in the mean-field $\hat\Gamma$ is
a $G$-matrix, the sum over all ladder diagrams. The
effective force in the $pp$ channel, i.e. in the pairing potential
$\hat\Delta$, should be the $K$ matrix, the
sum of all diagrams irreducible in $pp$-direction. However,
very little is known about this matrix in the relativistic
framework. And although the relativistic theory of pairing 
presents a very active area of research \cite{GCF.96,MFD.97}, 
only phenomenological effective forces have been shown
to produce reliable results when applied to finite nuclei, 
especially in exotic regions. In the present work we employ
a two-body finite range interaction of Gogny type \cite{BGG.84}
in the $pp$ channel of RHB: 
\begin{equation}
V^{pp}(1,2)~=~\sum_{i=1,2}
e^{-(( {\bf r}_1- {\bf r}_2)
/ {\mu_i} )^2}\,
(W_i~+~B_i P^\sigma 
-H_i P^\tau -
M_i P^\sigma P^\tau),
\end{equation}
with the parameters 
$\mu_i$, $W_i$, $B_i$, $H_i$ and $M_i$ $(i=1,2)$. 
The pairing interaction is a sum of two Gaussians 
with finite range and properly chosen spin and isospin dependence. 
The Gogny force has been very carefully adjusted to reproduce 
selected global properties of spherical nuclei and of 
nuclear matter. In the pairing channel its basic advantage 
is the finite range, which automatically guarantees a proper
cut-off in momentum space. This interaction was employed 
in the RHB calculations of Ref.~\cite{Gonz.96}.
For the D1S \cite{BGG.84} parameter set of the interaction
in the pairing channel, the model was applied in the study of several
isotope chains of spherical Pb, Sn and Zr nuclei.
In Refs.~\cite{PVL.97,LVP.98,LVR.98} we 
have used RHB in coordinate space with the D1S Gogny interaction
to describe properties of light nuclear systems (C, N, O, F, Ne, Na, Mg)
with large neutron excess, as well as ground-states of 
Ni ($28\leq N\leq 50$) and and Sn ($50\leq N\leq 82$) isotopes.

The eigensolutions of Eq. (\ref{equ.2.2}) form a set of
orthogonal and normalized single quasi-particle states. The corresponding
eigenvalues are the single quasi-particle energies.
The self-consistent iteration procedure is performed
in the basis of quasi-particle states. The resulting quasi-particle
eigenspectrum is then transformed into the canonical basis of single-particle
states, in which the RHB ground-state takes the  
BCS form. The transformation determines the energies
and occupation probabilities of the canonical states.

For nuclear systems with spherical symmetry the fields
$\sigma(r),\,\omega^0(r),\,\rho^0(r),$ and $A^0(r)$ depend
only on the radial coordinate $r$.  The nucleon spinors
$U_k$ ($V_k$) in (\ref{equ.2.2}) are characterized by the
angular momentum $j$, its $z$-projection $m$, parity $\pi$
and the isospin $t_3=\pm {1\over 2}$ for neutron and
proton. The two Dirac spinors $U_k({\bf r})$ and $V_k({\bf r})$
are defined
\begin{eqnarray}
\label{spherspinor}
{U_k(V_k)}({\bf r},s,t_3)=
\pmatrix{ g_{U(V)}(r)\Omega_{j,l,m} (\theta,\varphi,s) \cr
        if_{U(V)}(r)\Omega_{j,\tilde l,m} (\theta,\varphi,s) \cr }
         \chi_\tau(t_{3}).
\end{eqnarray}
$g(r)$ and $f(r)$ are radial amplitudes, $\chi_\tau$ is the
isospin function, and 
$\Omega_{jlm}$ is the tensor product of the orbital and
spin functions
\begin{equation}
\Omega_{j,l,m} (\theta,\varphi,s)=\sum\limits_{m_s,m_l}
\bigl< {1\over 2}m_slm_l\big\vert jm\bigr> 
\chi_{{1\over 2} m_s} Y_{lm_l}(\theta,\varphi).
\end{equation}
The two-component functions
\begin{equation}
\Phi_U(r):=\pmatrix{g_U(r)\cr i f_U(r)}~~~~~~{\rm and}~~~~~~ 
\Phi_V(r):=\pmatrix{g_V(r)\cr i f_V(r)} ,
\end{equation}
are solutions of the Dirac-Hartree-Bogoliubov equations 
\begin{eqnarray}
\label{equ..2.22}
(\hat h_D(r)-m -\lambda )\Phi_U(r)+
\int_0^{\infty}dr'r'^2\Delta(r,r')\Phi_V(r') 
= E\Phi_U(r) \nonumber \\
(-\hat h_D(r)+m +\lambda )\Phi_V(r)+
\int_0^{\infty}dr'r'^2\Delta(r,r')\Phi_U(r') 
= E\Phi_V(r)
\end{eqnarray}
The self-consistent solution of the 
Dirac-Hartree-Bogoliubov integro-differential eigenvalue equations
and Klein-Gordon equations for the meson fields determines the 
nuclear ground state. In Refs.~\cite{PVL.97,LVP.98,LVR.97,PVR2.97}
we have used Finite Element Methods
in the coordinate space discretization of the coupled system
of equations. Coordinate space solutions of the RHB equations
are essential for a correct description of nuclear structure phenomena
that originate from large spatial extensions of nucleon densities. These
include, for example, neutron skins and halos in very neutron-rich nuclei.
In less exotic nuclei on the neutron-rich side, or for proton-rich nuclei,
an expansion in a large oscillator basis should provide sufficiently 
accurate solutions \cite{DFT.84,DNW.96}. In particular, proton-rich nuclei are 
stabilized by the Coulomb barrier which tends to localize the proton
density in the nuclear interior and thus prevents the formation
of objects with extreme spatial extension. In the present work 
we employ the procedure of Refs.~\cite{Gonz.96,LVR.98}, and solve
the Dirac-Hartree-Bogoliubov equations and the equations for the 
meson fields  by expanding the nucleon spinors 
$U_k({\bf r})$ and $V_k({\bf r})$, 
and the meson fields in a basis of spherical harmonic oscillators 
for $N = 20$ oscillator shells~\cite{GRT.90}.  However, in order to 
verify that our final conclusions do not depend on the method of solution,
for nuclei at the proton drip-line we have also performed RHB calculations
in coordinate space ~\cite{PVR2.97}. In particular, 
coordinate space solutions have confirmed
our predictions for the location of the proton-drip line.
%
%
%
\section {Ground-state properties of proton-rich nuclei}
In the present application of the Relativistic Hartree Bogoliubov theory
we describe the ground-state properties of spherical even-even nuclei
$14\leq Z \leq 28$ and $N=18,20,22$. While for these neutron
numbers the nuclei with 
$14\leq Z \leq 20$ are not really very proton-rich, 
nevertheless they will be useful for 
a comparison of the model calculations with experimental data. We are 
particularly interested in the predictions of the model for 
the proton-rich nuclei in the 1f$_{7/2}$ region. These nuclei 
have recently been extensively investigated in experiments
involving fragmentation of $^{58}$Ni \cite{Det.90,Bor.92,Bla.94,Bla.96}. 
The principal motivation of many experimental studies in this region is
the possible occurrence of the two-proton ground-state 
radioactivity. In particular, the region around $^{48}$Ni is expected to 
contain nuclei which are two-proton emitters. On the other hand, 
because of the Coulomb barrier at the proton drip-line, 
the emission of a pair of protons may be strongly delayed for 
nuclei with small negative two-proton separation energies.

The input for our calculations are the coupling constants and 
masses for the effective mean-field Lagrangian, and the 
effective interaction in the pairing channel.
In the analysis of light neutron-rich nuclei in 
Refs. \cite{PVL.97,LVP.98,LVR.97}, as well as in the study of ground-state
properties of Ni and Sn isotopes \cite{LVR.98}, we have used the NL3 
parameter set for the effective mean-field Lagrangian in the $ph$ channel.
The effective interaction  NL3 has been derived 
\cite{LKR.97} by adjusting model calculations to bulk properties
of a large number of spherical nuclei.  Properties
calculated with the NL3 effective interaction are found to
be in very good agreement with experimental data for nuclei
at and away from the line of $\beta$-stability. 
In Ref. \cite{LR.98} it has been shown that constrained 
RMF calculations with the NL3 effective force reproduce the 
excitation energies of superdeformed minima relative to the 
ground-state in $^{194}$Hg and $^{194}$Pb. In the same work 
the NL3 interaction was also used for calculations 
of binding energies and deformation parameters of rare-earth
nuclei. In the present study for the first time we employ the 
NL3 effective force on the proton-rich side of
the $\beta$-stability line. In view of the fact that all the results 
obtained so far indicate that NL3 is probably the best effective
RMF interaction, the main purpose of the analysis is to study 
how well  the properties predicted by the NL3 force compare 
with experimental data for proton-rich nuclei. However, 
in order to be more specific in our predictions for 
the exact location of the proton drip-line, we will also use two 
additional standard RMF effective interactions:
NL1~\cite{RRM.86} and NL-SH~\cite{SNR.93}. 
These effective forces have been used in many analyses to
calculate properties of nuclear matter and of finite nuclei, 
and generally produce very good results for nuclei close 
to the $\beta$-stability line.
In particular, the effective interaction NL1 was also
used in the RHB+Gogny calculations of Ref. \cite{Gonz.96}
 
In most applications of relativistic mean-field theory 
pairing correlations have been included in the form 
of a simple BCS approximation, with a monopole pairing force
adjusted to the experimental odd-even mass differences \cite{GRT.90}.
For nuclei far from the valley of $\beta$-stability
this approach becomes unreliable, especially in the 
calculation of properties that crucially depend on
the spatial extensions of nucleon densities. The BCS 
description of the scattering of nucleonic pairs from
bound states to the positive energy particle continuum
produces an unphysical component in the nucleon density
with the wrong asymptotic behavior~\cite{DFT.84,DNW.96}.
This effect is more pronounced for neutron-rich nuclei,
for which the coupling to the particle continuum is 
particularly important. For proton-rich nuclei the 
Coulomb barrier confines the protons in the interior 
of the nucleus, and therefore the effect of the 
coupling to the continuum is weaker. However, if pairing 
correlations are described in the unified framework of the 
RHB scheme (or HFB in the non-relativistic approach), 
the nucleon densities display a correct asymptotic 
behavior. The effective interactions that have been 
used in the pairing channel of RHB are the pairing 
part of the Gogny force and the density-dependent
delta force. The finite range interaction provides 
an automatic cut-off of high momentum components, 
while an artificial energy cut-off has to be included
in the calculation with zero-range forces. On the 
other hand, the density-dependent interaction 
can be adjusted to produce surface peaked pairing 
fields, which can be important for a correct description 
of spatial distribution of densities.  A fully self-consistent 
RHB model in coordinate space, with a density dependent
interaction of zero-range (delta force), has been used
to describe the two-neutron halo in $^{11}$Li~\cite{MR.96}.
In the present study we employ the pairing part of the 
Gogny interaction in the $pp$ channel, 
with the parameter set D1S \cite{BGG.84}. 

In Fig. \ref{figA} we display the two-proton separation 
energies
\begin{equation}
S_{2p}(Z,N) = B_p(Z,N) - B_p(Z-2,N)
\label{sep}
\end{equation}
for the even-even nuclei $14\leq Z \leq 28$ and $N=18,20,22$.
The values that correspond to the self-consistent RHB ground-states 
are compared with experimental data and extrapolated values from 
Ref. \cite{AW.95}. We notice that the theoretical values 
reproduce in detail the experimental separation energies,
except for $^{38}$Ca and $^{44}$Ti. In order to understand better
this result, in Table \ref{TabA} we compare the calculated
total binding energies for the $N=18,20,22$  
isotones with empirical values. We find that our model 
results are in very good agreement with experimental data 
when one of the shells (proton or neutron) is closed, or 
when valence protons and neutrons occupy different major 
shells (i.e. bellow and above $N$ and/or $Z=20$). 
The absolute differences between the calculated and experimental 
masses are less than 2 MeV. 
The differences are larger when both proton and neutron valence
particles (holes) occupy the same major shell, and especially 
for the $N=Z$ nuclei $^{36}$Ar and $^{44}$Ti. This seems to
be a clear indication that for these nuclei additional
correlations should be taken into account.
In particular, proton-neutron pairing 
could have a strong influence on the masses. Proton-neutron 
short-range correlations are not included in our model.

The results should be also compared with recently reported 
self-consistent mean-field calculations of Ref. \cite{Naz.96},
and with properties of proton-rich nuclei calculated within
the framework of the nuclear shell model \cite{Orm.96}.
The calculations of  Ref. \cite{Naz.96} have been performed 
for several mean-field models
(Hartree-Fock, Hartree-Fock-Bogoliubov, and relativistic 
mean-field), and for a number of effective interactions.
The results systematically 
predict the two-proton drip-line to lie between 
$^{42}$Cr and $^{44}$Cr, $^{44}$Fe and $^{46}$Fe, 
and $^{48}$Ni and $^{50}$Ni. Very recent studies of
proton drip-line nuclei in this region have been
performed in experiments based on $^{58}$Ni
fragmentation on a beryllium target \cite{Bla.94,Bla.96}.
In Ref. \cite{Bla.94} in particular, evidence has 
been reported for particle stability of $^{50}$Ni.
In the shell-model calculations of Ref. \cite{Orm.96}
absolute binding energies were evaluated by computing 
the Coulomb energy shifts between mirror nuclei, and 
adding this shift to the experimentally determined binding 
energy of the neutron rich isotope. The calculated two-proton
separation energies predicted a proton drip-line in agreement
with experimental data and
with the mean-field results \cite{Naz.96}. Compared to the 
results of the present study, the shell-model total binding 
energies are in somewhat better agreement with experimental data.
However, the two models give almost identical values for 
the extracted two-proton separation energies of the 
drip-line nuclei. The self-consistent 
RHB NL3+D1S two-proton separation energies
at the drip-line are also very close to the values 
that result from non-relativistic HFB+Gogny (D1S) calculation 
of Ref. \cite{Naz.96}.

By using a fully microscopic and 
self-consistent model for the calculation of binding energies,
we have the possibility to analyze in detail the single-proton  
levels. In Figs. \ref{figB}, \ref{figC}, \ref{figD} 
we display the proton single-particle 
energies in the canonical basis as functions of proton
number for the $N=18,20,22$ isotones, respectively. The thick solid
lines denote the corresponding Fermi levels. Although the 
proton levels do not change much with $Z$, we observe a
consistent decrease in the energy splitting between 
the spin-orbit partners 1d$_{5/2}$ - 1d$_{3/2}$ 
and 2p$_{3/2}$ - 2p$_{1/2}$ with increasing proton number.
We will show that this decrease results from the reduction 
of the spin-orbit term of the effective potential \cite{LVR.97}.
The 1f$_{7/2}$ orbital is unbound for all  $N=18$ isotones,
and is very slightly bound for $N=20$.
The Fermi level displays a sharp increase with $Z$ for 
all three isotone chains. In principle, a positive 
value of $\lambda$ should indicate which 
nuclei are beyond the proton drip-line, i.e. which nuclei
are ground-state proton emitters. In particular, for
$^{42}$Cr, $^{46}$Fe and $^{50}$Ni we find $\lambda > 0$.
This is somewhat surprising, since for $^{46}$Fe and $^{50}$Ni
the calculated two-proton separation energies are positive.
We have performed RHB calculations with the effective
interactions NL1 and NLSH, but also for these forces 
the Fermi level is positive for $^{42}$Cr, $^{46}$Fe 
and $^{50}$Ni. For these three nuclei we have also 
verified the results by performing coordinate space RHB
calculations. The results are practically identical to those 
obtained with the oscillator expansion method; the 
Fermi levels for these three nuclei have positive values. Therefore
it appears that there are cases at the drip-line for which the 
definition of the two-particle separation energy (\ref{sep}) 
does not correspond to the physical interpretation of the chemical
potential.

In Fig. \ref{figE} we show the self-consistent ground-state
proton densities for the $N=20$ isotones. The 
density profiles display shell effects in the bulk and 
a gradual increase of proton radii. In the
insert of Fig. \ref{figE} we include the corresponding values for
the surface thickness and diffuseness parameter. The
surface thickness $t$ is defined to be the change in radius
required to reduce $\rho (r) / \rho_0$ from 0.9 to 0.1
($\rho_0$ is the maximal value of the neutron density; because of
shell effects we could not use for  $\rho_0$ the density
in the center of the nucleus). The
diffuseness parameter $\alpha$ is determined by fitting the
neutron density profiles to the Fermi distribution
\begin{equation}
\rho (r) =  {\rho_0} \left (1 + exp({{r - R_0}\over 
\alpha})\right)^{-1} ,
\end{equation}
where $R_0$ is the half-density radius. In going away from the
valley of $\beta$-stable nuclei, generally  
the proton surface thickness increases
and the surface becomes more diffuse. However, while $t$  
increases from Si to Ni, the diffuseness parameter
$\alpha$ has a maximum at $Z=20$. It appears that, as protons
fill the 1f$_{7/2}$ orbital, the proton surface becomes 
slightly less diffuse. This could be due to the stronger
influence of the Coulomb barrier. In \ref{figF} we display 
the self-consistent proton potentials for the $N=20$ isotones,
and in the insert the details of the potentials in the 
region of the Coulomb barrier. We notice how the 
Coulomb barrier increases from 3 MeV for $^{34}$Si, to
6 MeV in $^{48}$Ni. We include also $^{48}$Ni
in our figures for the $N=20$ isotones, although this 
nucleus is not particle stable in our calculations.

In Fig. \ref{figG} we display the 
proton $rms$ radii for $N=18,20,22$ isotones, respectively. The 
calculated values are compared with experimental 
data for proton radii from Ref. \cite{VJV.87}.
Except for $^{32}$Si, we find an excellent agreement
between experimental data and values calculated with the 
NL3 effective force with the D1S Gogny interaction in the 
pairing channel. The model predicts a uniform increase 
of $rms$ radii with the number of protons. 

In an analysis of ground-state properties of light 
neutron-rich nuclei \cite{LVR.97}, 
we have shown that the relativistic
mean-field model predicts a strong isospin dependence 
of the effective spin-orbit potential. With the increase
of the number of neutrons the effective spin-orbit interaction
becomes weaker and the magnitude of the spin-orbit term in the 
single nucleon potential is significantly reduced. 
This results in a reduction 
of the energy splittings for spin-orbit partners. The reduction in the 
surface region was found to be as large as $\approx 40 \%$ for Ne 
isotopes at the drip-line. In Ref. \cite{LVR.98} similar 
results were found for the Ni and Sn isotopes. The spin-orbit
potential originates from the addition of two large fields:
the field of the vector mesons (short range repulsion), and
the scalar field of the sigma meson (intermediate
attraction). In the first order approximation, and
assuming spherical symmetry, the spin orbit term can be
written as
\begin{equation}
\label{so1}
V_{s.o.} = {1 \over r} {\partial \over \partial r} V_{ls}(r), 
\end{equation} 
where $V_{ls}$ is the spin-orbit potential~\cite{KR2.91}
\begin{equation}
\label{so2}
V_{ls} = {m \over m_{eff}} (V-S).
\end{equation}
V and S denote the repulsive vector and the attractive
scalar potentials, respectively.  $m_{eff}$ is the
effective mass
\begin{equation}
\label{so3}
m_{eff} = m - {1 \over 2} (V-S).
\end{equation}
Using the vector and scalar potentials from the NL3
self-consistent ground-state solutions, we have computed
from~(\ref{so1}) - (\ref{so3}) the spin-orbit terms for the
$N=20$ isotones.  They are shown in Fig. \ref{figH} as function of
the radial distance from the center of the nucleus. 
The magnitude of the spin-orbit term $V_{s.o.}$ decreases
as we add more protons, i.e. as we move away from $\beta$-stable nuclei.
>From $^{34}$Si to $^{48}$Ni, 
the reduction is $\approx 20\%$ in the surface region.
The minimum of $V_{s.o.}$ is also shifted outwards, and this  
reflects the larger spatial extension of the proton densities.
However, we note that the reduction of $V_{s.o.}$ for protons
is considerably smaller than the one calculated for neutrons in 
Refs. \cite{LVR.97,LVR.98} ($\approx 35-40\%$). 

The properties of the finite-range and density independent
pairing interaction are illustrated in 
Figs. \ref{figI} and \ref{figJ}. In Fig. \ref{figI} 
we plot the average values 
of the proton canonical pairing gaps $\Delta_{nlj}$ 
as functions of canonical single-particle energies. The gaps 
are displayed for canonical states that 
correspond to the self-consistent ground state of $^{44}$Cr.
$\Delta_{nlj}$ are the diagonal matrix elements
of the pairing part of the RHB single-nucleon Hamiltonian in the 
canonical basis. Although not completely equivalent, $\Delta_{nlj}$
corresponds to the pairing gap in BCS theory. A very detailed
discussion of HFB equations in the canonical basis can be found in
Ref. \cite{DNW.96}.
The pairing gaps have relatively large values for deep-hole states.
This is related to the volume character of the Gogny interaction in the 
pairing channel. The average value at the Fermi surface is
between 1.5 and 2 MeV, and $\Delta_{nlj}$ slowly decrease 
for canonical states in the single-proton continuum.
In Fig. \ref{figJ} we display the averages of the proton  
pairing gaps for occupied canonical states
\begin{equation}
< \Delta_p > = {{\sum_{nlj} \Delta_{nlj} v_{nlj}^2}\over
                  {\sum_{nlj}  v_{nlj}^2}},
\label{ang}
\end{equation}
where $v_{nlj}^2$ are the occupation probabilities. 
The values of $< \Delta_p >$ for the $N=22$ isotones are
plotted as function of the proton number. 
The average proton gap increases to almost 3 MeV for 
$^{38}$Ar, then the pairing correlations disappear
at shell closure $Z=20$.
For the 1f$_{7/2}$ orbital the value of $< \Delta_p >$
is $\approx$ 2.5 MeV.

In conclusion, this study presents the first application
of the Relativistic Hartree
Bogoliubov theory to the description of ground-state properties
of proton-rich nuclei. A detailed analysis 
of spherical even-even nuclei with $14\leq Z \leq 28$ and $N=18,20,22$
has been performed.
The NL3 parameter set has been used for the effective 
mean-field Lagrangian in the $ph$ channel,
and pairing correlations have been described
by the finite range Gogny interaction D1S. 
In a comparison with available experimental 
data it has been shown that the NL3 + Gogny D1S effective
interaction provides a very good description of binding energies, 
two-proton separation energies and proton $rms$ radii.
Model predictions for the proton drip-line agree with recently
reported calculations in the framework of the nuclear shell-model 
and with results of non-relativistic HF and HFB studies.
For isotone chains we have also discussed the predicted
reduction of the effective spin-orbit potential with the 
increase of the number of protons, as well as the resulting 
energy splittings between spin-orbit partners and 
modifications of surface properties.

\begin{table}
\caption{ Comparison between calculated and empirical
binding energies. All values are in units of MeV; empirical 
values are displayed in parentheses.}
\begin{center}
\begin{tabular}{llllll}
\hline  
$^{32}$Si& 269.02 (271.41)&$^{40}$Ar&343.97 (343.81)&$^{44}$Cr&351.65 (349.99)\\
$^{34}$Si& 284.42 (283.43)&$^{38}$Ca&313.11 (313.04)&$^{46}$Cr&380.19 (381.98)\\
$^{36}$Si& 293.08 (292.02)&$^{40}$Ca&341.99 (342.05)&$^{44}$Fe&312.07 (-)     \\
$^{34}$S & 288.10 (291.84)&$^{42}$Ca&362.95 (361.90)&$^{46}$Fe&352.25 (350.20)\\
$^{36}$S & 307.98 (308.71)&$^{40}$Ti&315.39 (314.49)&$^{48}$Fe&384.42 (385.19)\\
$^{38}$S & 320.77 (321.05)&$^{42}$Ti&348.35 (346.91)&$^{46}$Ni&306.72 (-)     \\
$^{36}$Ar&302.52 (306.71) &$^{44}$Ti&373.15 (375.47)&$^{48}$Ni&349.92 (-)     \\
$^{38}$Ar&327.34 (327.06) &$^{42}$Cr&314.94 (314.20)&$^{50}$Ni&385.52 (385.50)\\
\end{tabular}
\end{center}
\label{TabA}
\end{table}

\begin{figure}
\caption{ Comparison between RHB/NL3 and experimental 
two-proton separation energies for $N=18,20,22$ isotones. 
Black symbols denote empirical values; lines connect 
symbols which correspond to calculated values.}
\label{figA}
\end{figure}

\begin{figure}
\caption{The proton single-particle levels
for the $N=18$ isotones. Solid lines denote the
neutron Fermi level. The energies in the canonical basis 
correspond to ground-state solutions calculated with the
NL3 effective force of the mean-field
Lagrangian. The parameter set D1S is used for
the finite range Gogny-type interaction in the
pairing channel.}
\label{figB}
\end{figure}

\begin{figure}
\caption{Same as in Fig. 2, but for the $N=20$ isotones.}
\label{figC}
\end{figure}

\begin{figure}
\caption{Same as in Fig. 2, but for the $N=22$ isotones.}
\label{figD}
\end{figure}

\begin{figure}
\caption{ Self-consistent RHB single-proton density 
distributions for the $N=20$ isotones, calculated with the
NL3 effective interaction.}
\label{figE}
\end{figure}

\begin{figure}
\caption{Self-consistent proton potentials for the $N=20$ isotones.
In the insert the details of the Coulomb barriers are shown.}
\label{figF}
\end{figure}

\begin{figure}
\caption{Calculated proton $rms$ radii for $N=18,20,22$ isotones
compared with experimental data.}
\label{figG}
\end{figure}

\begin{figure}
\caption{Radial dependence of the spin-orbit term
of the potential in self-consistent solutions for  
ground-states of the $N=20$ isotones.}
\label{figH}
\end{figure}

\begin{figure}
\caption{Average values of the proton canonical pairing gaps 
as functions of canonical single-particle energies for states that 
correspond to the self-consistent ground state of $^{44}$Cr. The NL3 
parametrization has been used for the mean-field
Lagrangian, and the parameter set D1S for the 
pairing interaction.}
\label{figI}
\end{figure}

\begin{figure}
\caption{ Average proton pairing gaps $< \Delta_p >$ 
of the $N=22$ isotones.}
\label{figJ}
\end{figure}

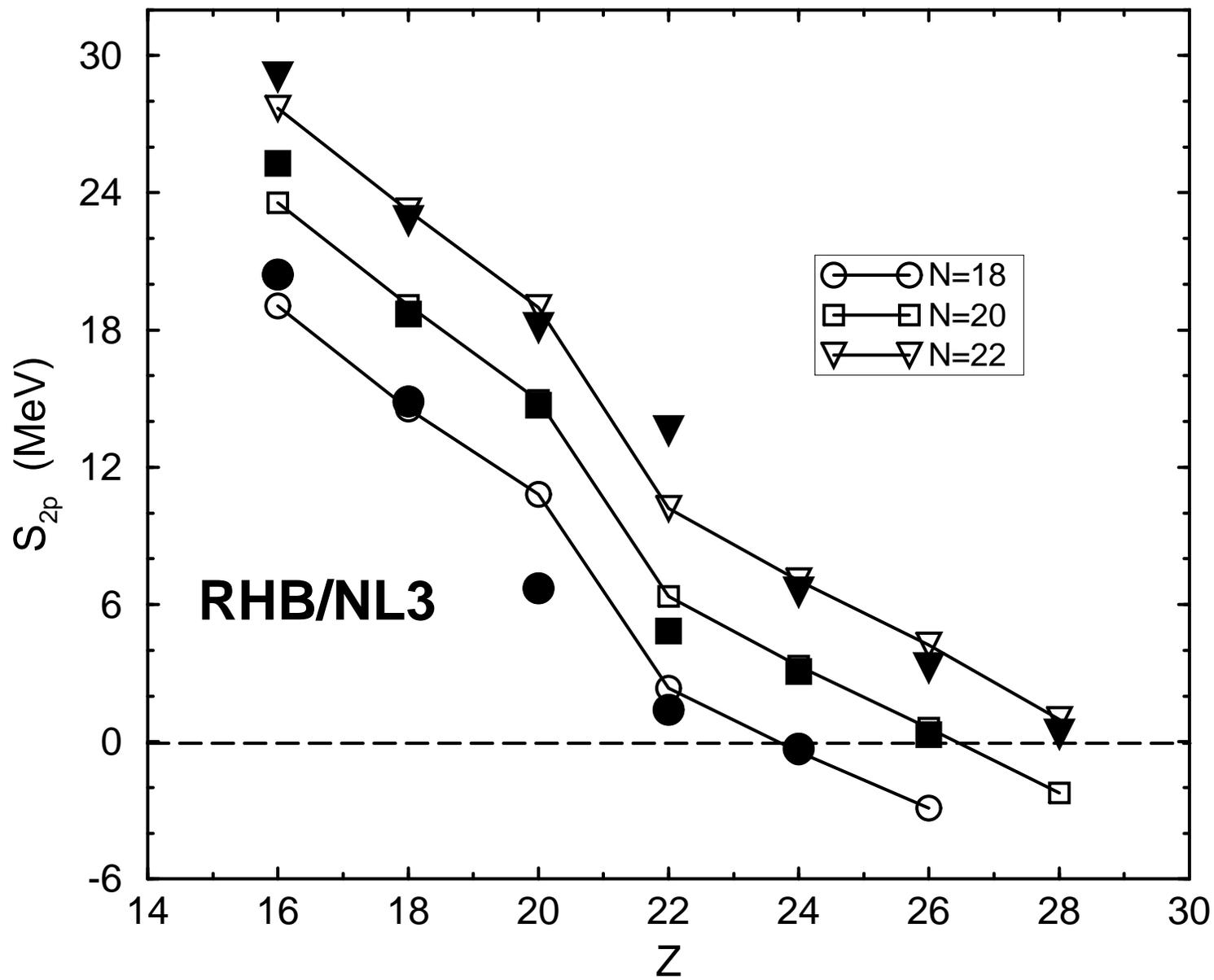
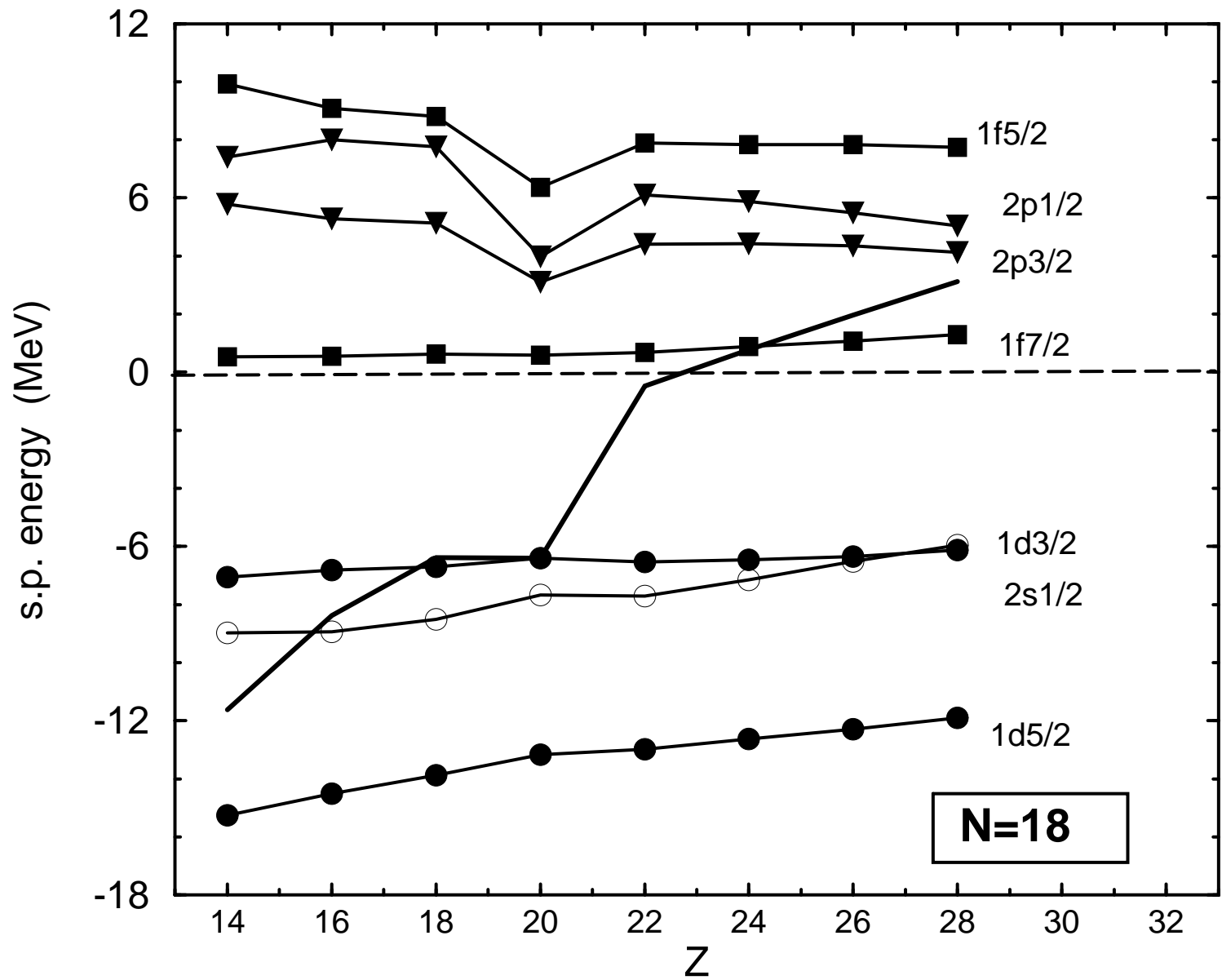
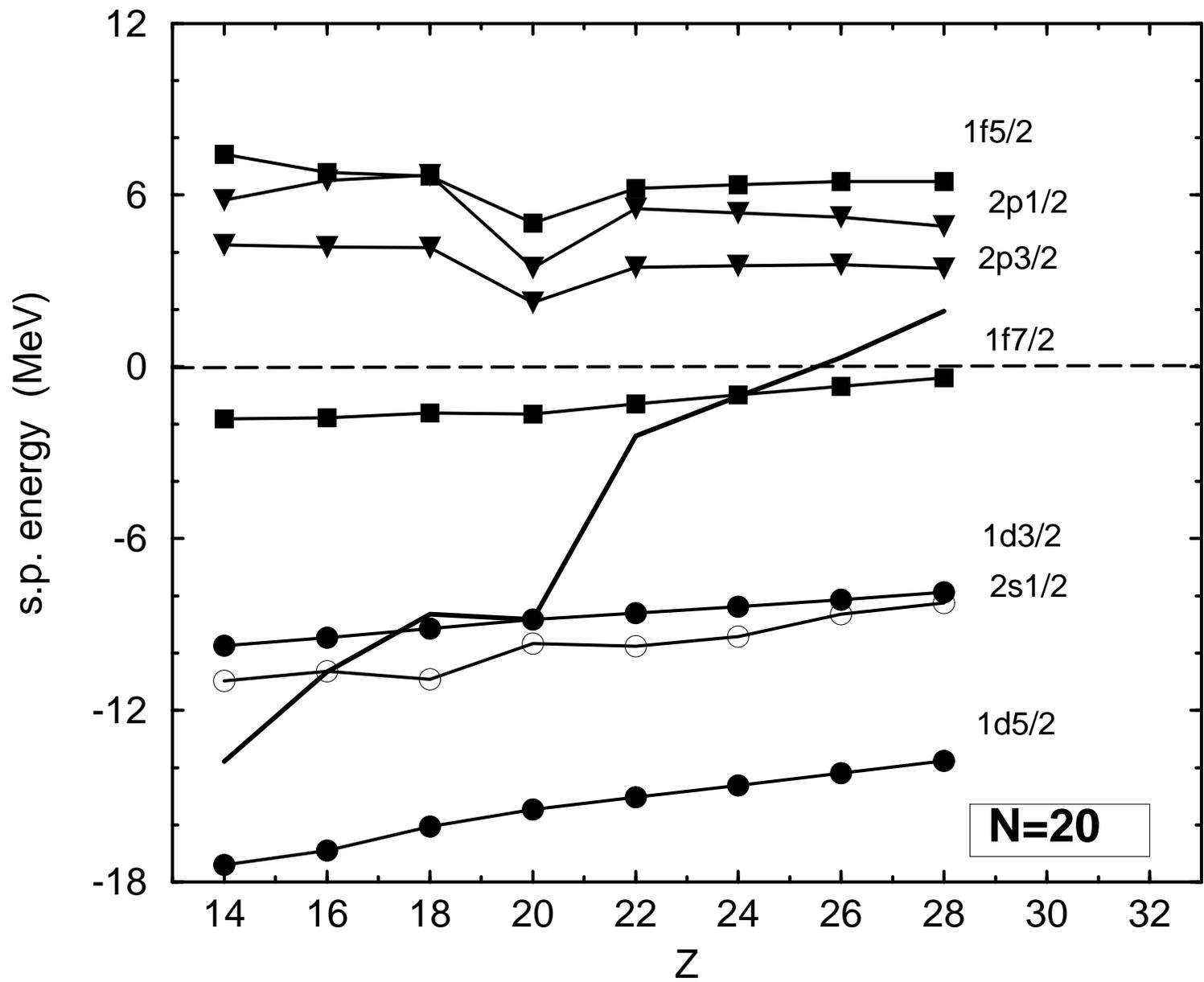
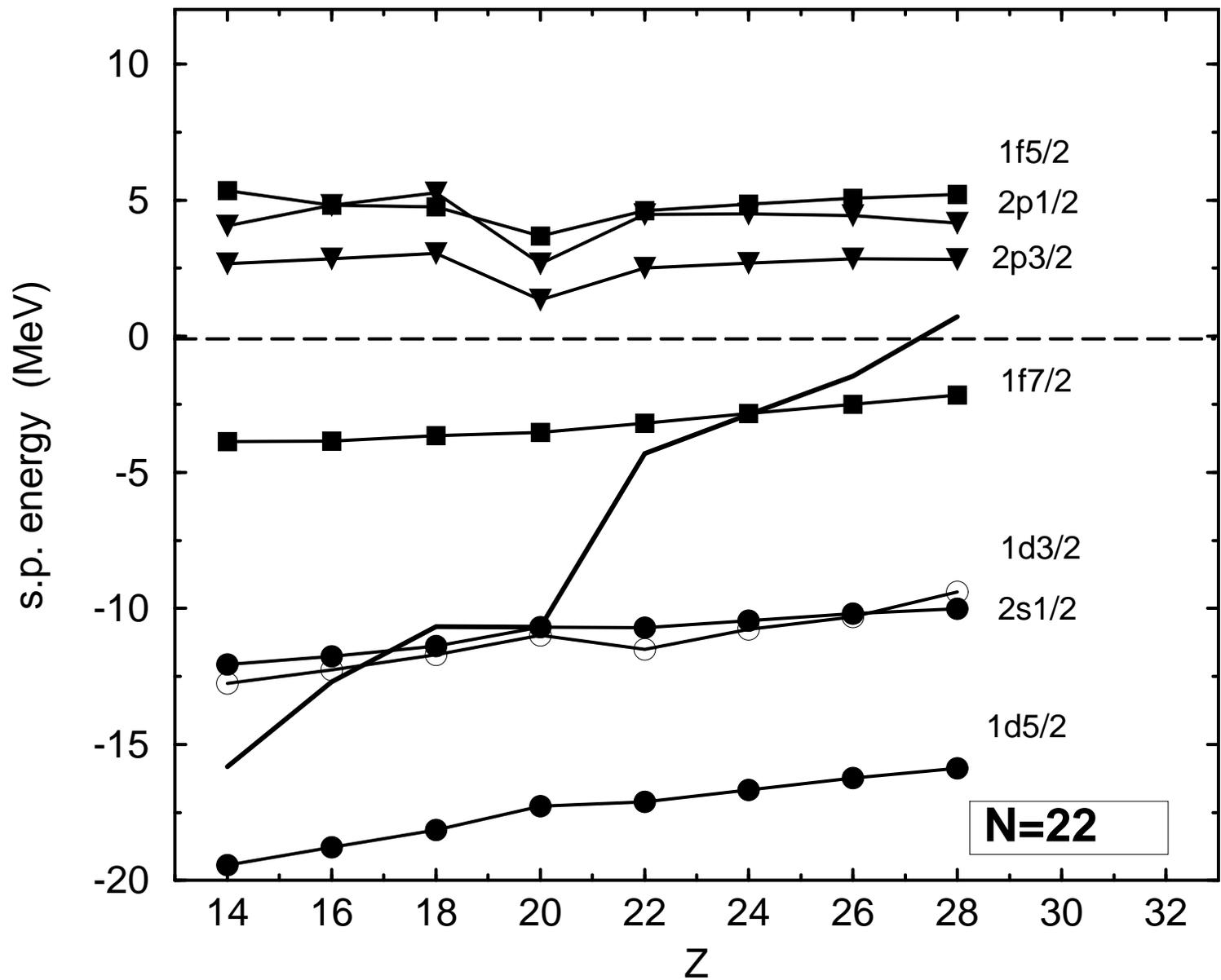
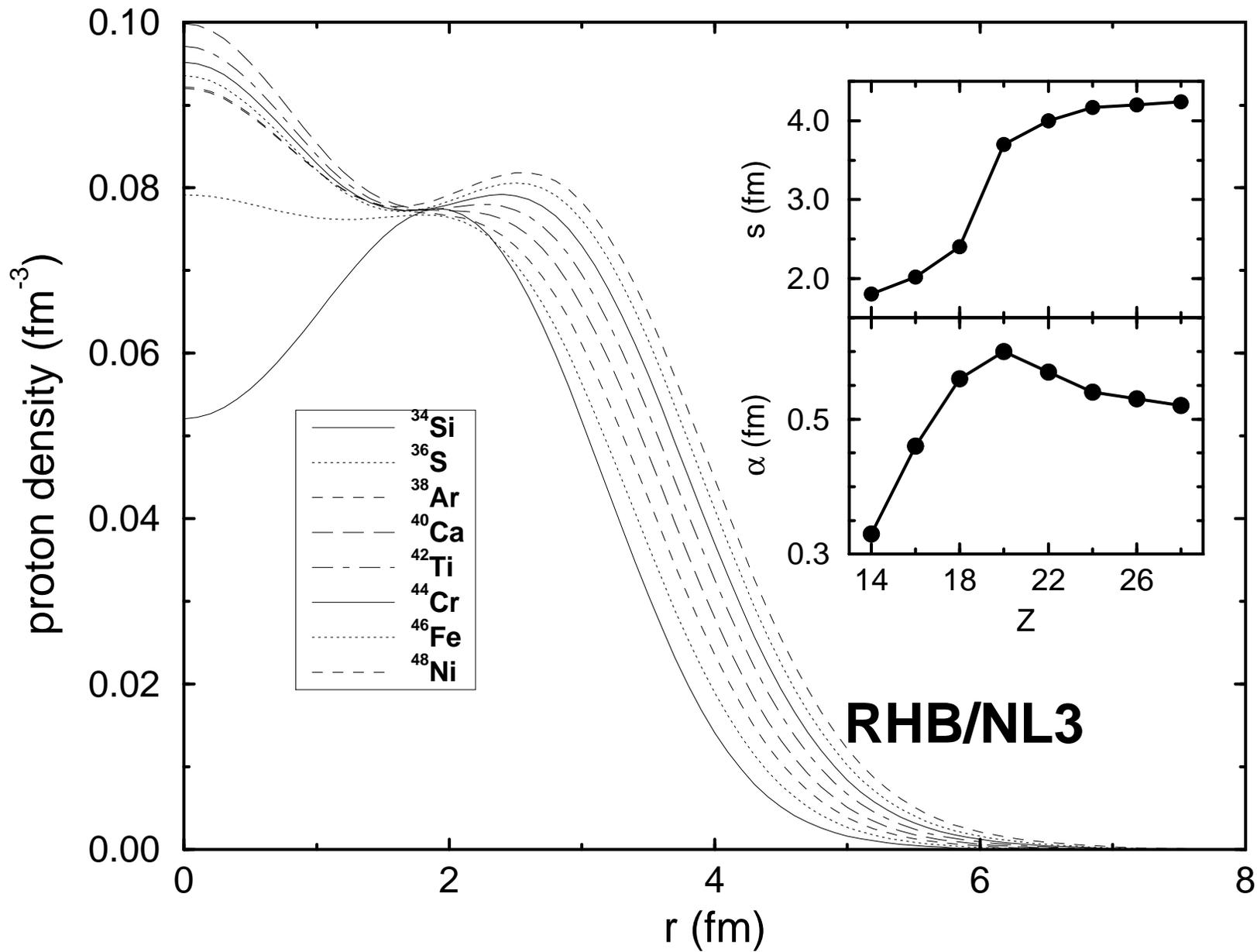
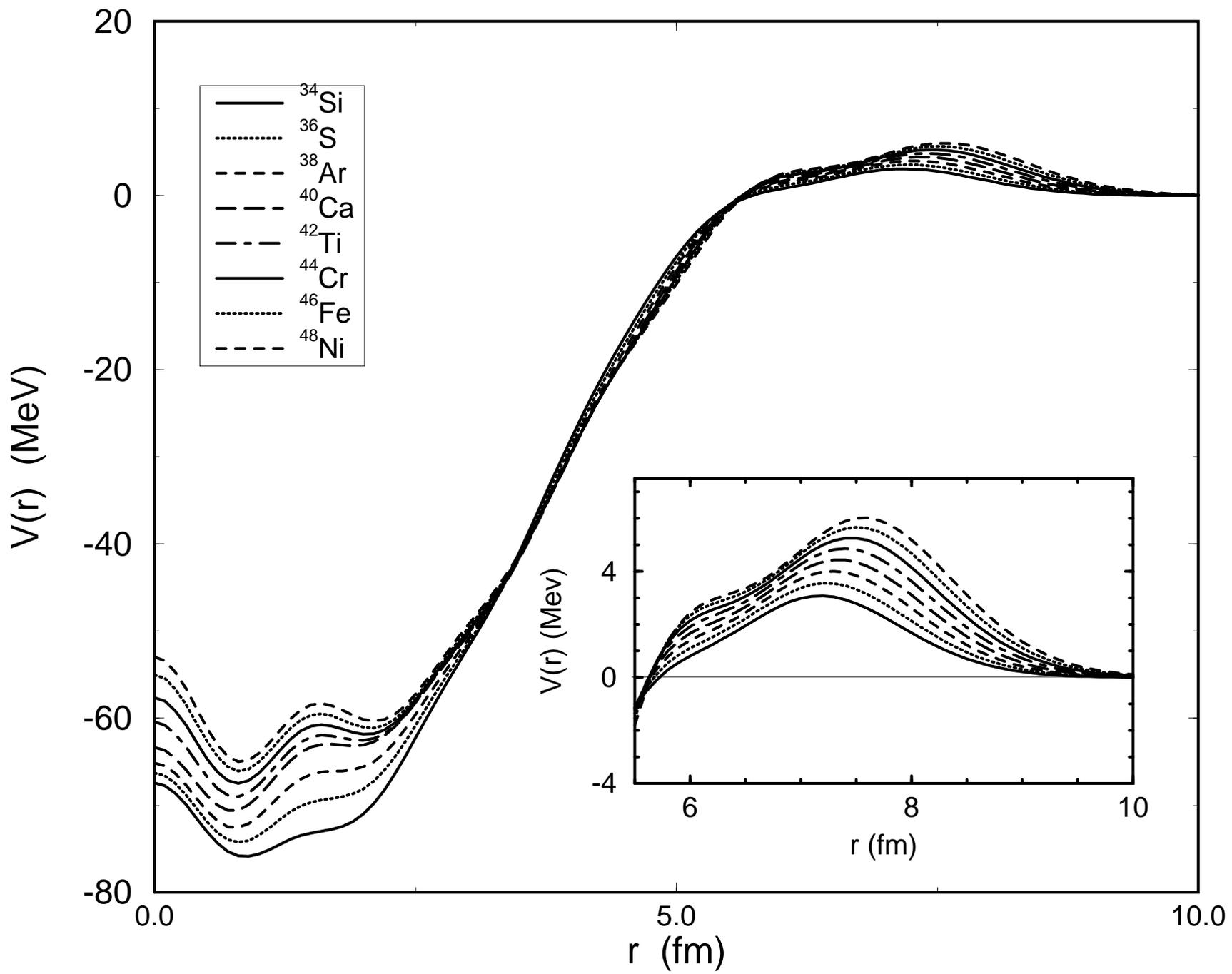
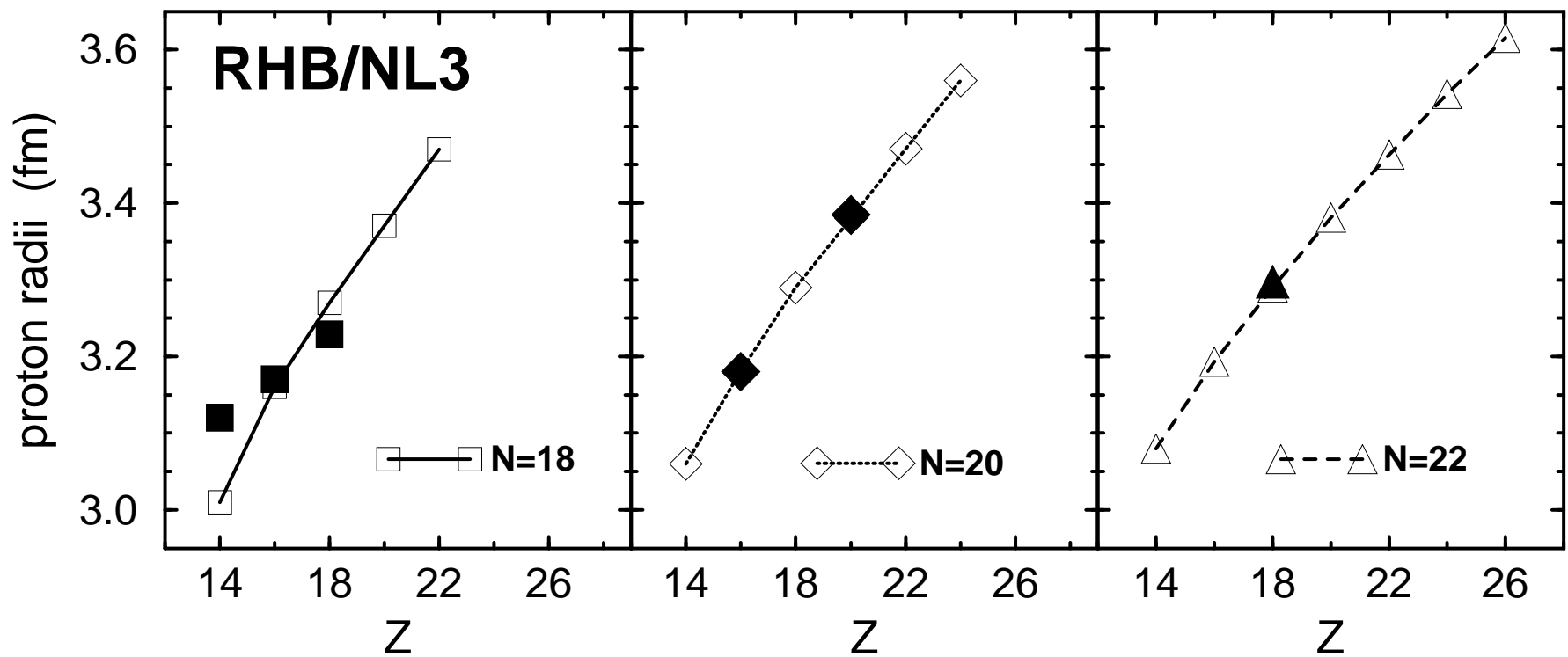
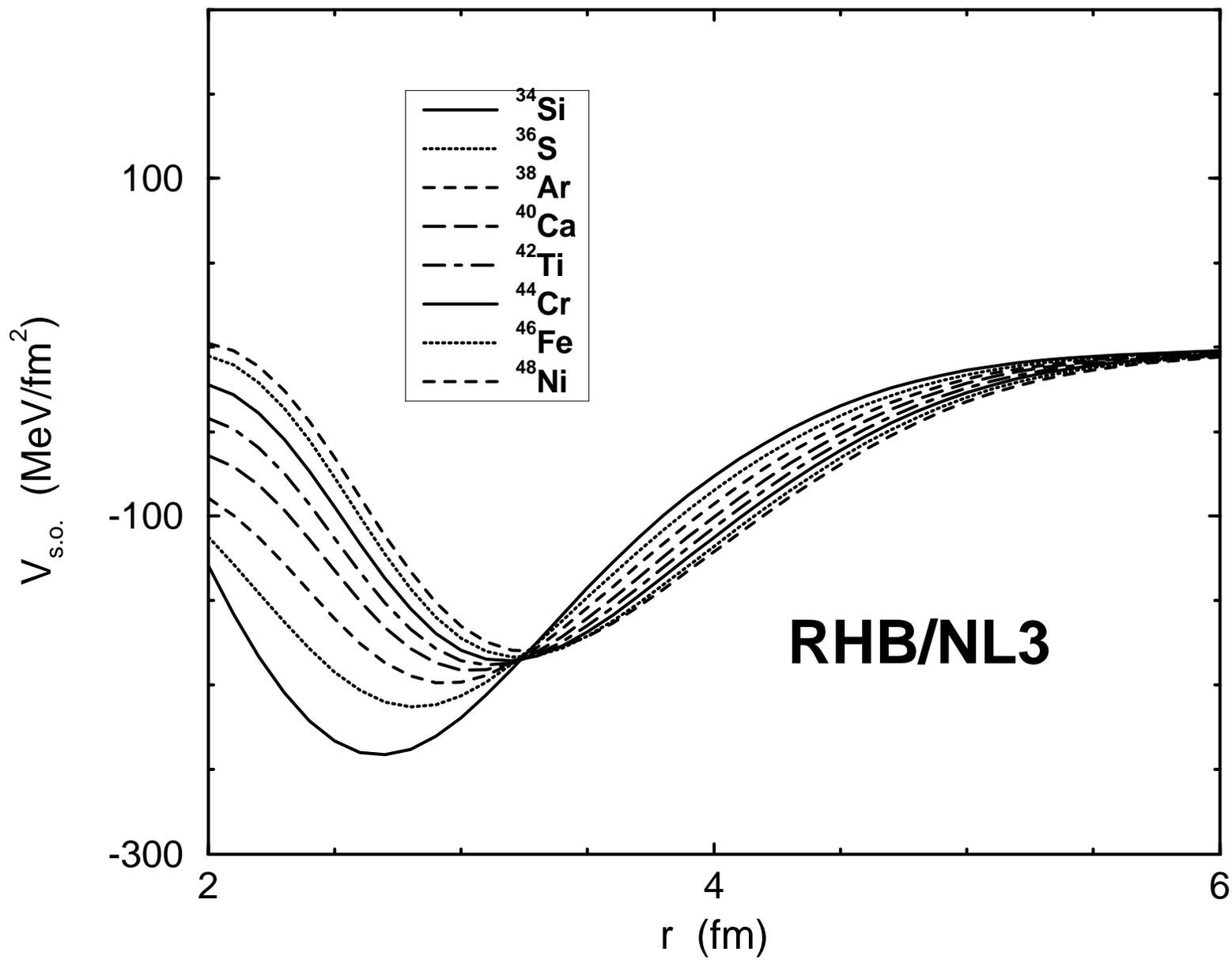
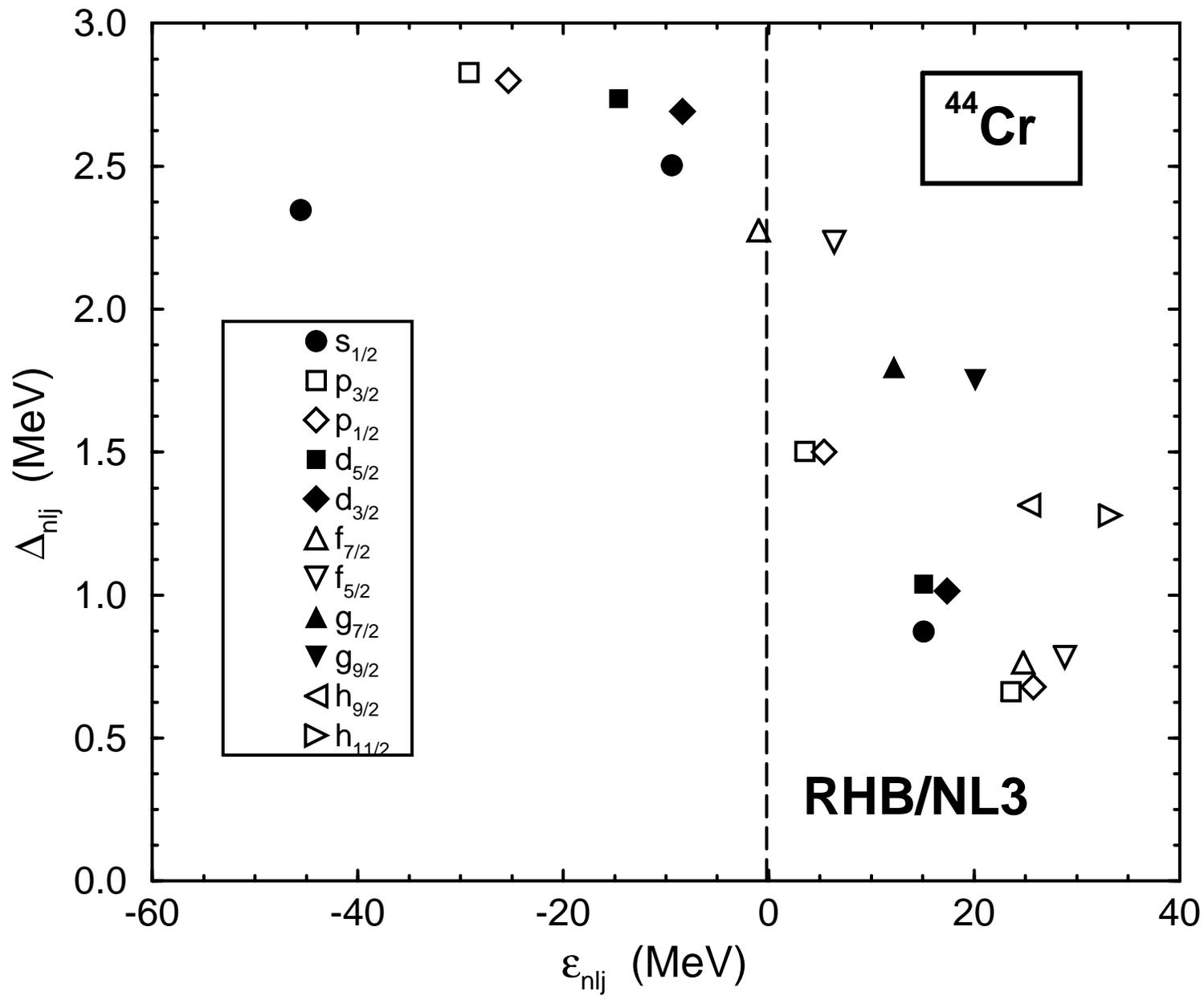
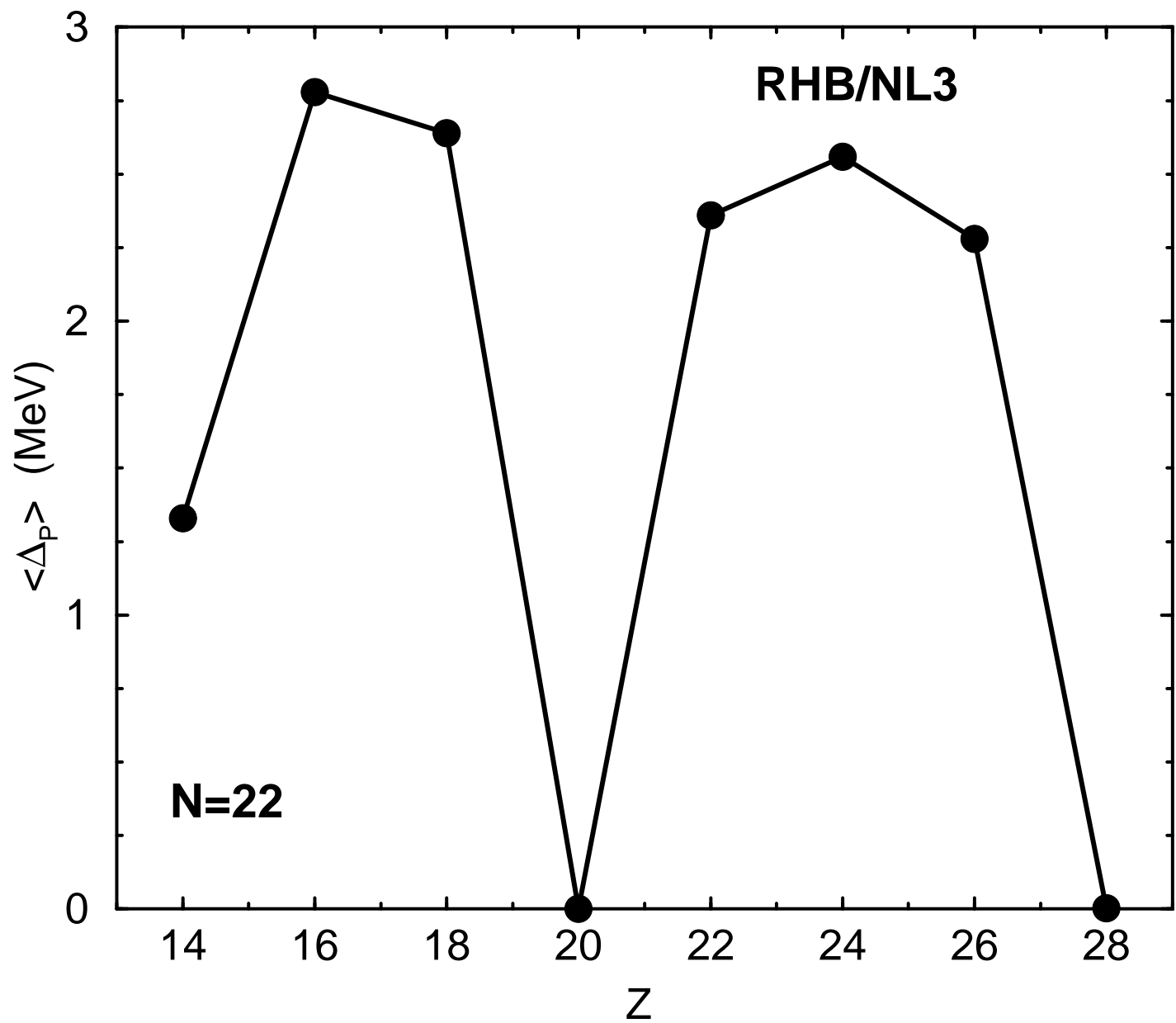
\end{document}